\documentclass[a4paper,fleqn,usenatbib]{mnras}

\usepackage{mathptmx}

\usepackage[T1]{fontenc}
\usepackage{ae,aecompl}

\usepackage{graphicx}	
\usepackage{amsmath}	
\usepackage{amssymb}	
\usepackage{color}

\def\simlt{\mathrel{\hbox{\rlap{\hbox{\lower4pt\hbox{$\sim$}}}\hbox{$<$}}}}
\def\simgt{\mathrel{\hbox{\rlap{\hbox{\lower4pt\hbox{$\sim$}}}\hbox{$>$}}}}

\newcommand{\tmerge}{t_{\textrm{merge}}}
\newcommand{\twr}{t_{\textrm{WR}}}
\newcommand{\dmerge}{d_{\textrm{merge}}}
\newcommand{\twind}{t_{\textrm{wind}}}
\newcommand{\ttau}{t_{\tau}}
\newcommand{\dtau}{d_{\tau}}

\newcommand{\async}{a_{\textrm{sync}}}

\newcommand{\chieff}{\chi_{\textrm{eff}}}

\title{Research Note:  The Expected Spins of {Gravitational Wave Sources With Isolated Field Binary Progenitors}}

\author[M. Zaldarriaga et al.]{
Matias Zaldarriaga,$^{1}$\thanks{E-mail: matiasz@ias.edu}
Doron Kushnir,$^{2}$
Juna A. Kollmeier$^{3}$
\\
$^{1}$School of Natural Sciences, Institute for
Advanced Study, Princeton, NJ, 08540, USA\\
$^{2}$Dept. of Particle Phys. \& Astrophys., Weizmann Institute of
Science, Rehovot 76100, Israel\\
$^{3}$ Observatories of the Carnegie Institution of Washington,
  813 Santa Barbara Street, Pasadena, CA 91101, USA
}

\date{Accepted XXX. Received YYY; in original form ZZZ}

\pubyear{2017}

\begin{document}
\label{firstpage}
\pagerange{\pageref{firstpage}--\pageref{lastpage}}
\maketitle

\begin{abstract}

We explore the consequences of dynamical evolution of field binaries composed of a primary black hole (BH) and a Wolf-Rayet (WR) star in the context of gravitational wave (GW) source progenitors. We argue, from general considerations, that the spin of the WR-descendent BH will be maximal in a significant number of cases due to dynamical effects. In other cases, the spin should reflect the natal spin of the primary BH which is currently theoretically unconstrained. We argue that the three currently published LIGO systems (GW150914, GW151226, LVT151012) suggest that this spin is small.  The resultant effective spin distribution of gravitational wave sources should thus be bi-model if this classic GW progenitor channel is indeed dominant. While this is consistent with the LIGO detections thus far, it is in contrast to the three best-measured high-mass x-ray binary (HMXB) systems. A comparison of the spin distribution of HMXBs and GW sources should ultimately reveal whether or not these systems arise from similar astrophysical channels.

\end{abstract}



\section{Introduction}


In this brief note we follow the set up described in \citep{Kushnir2016MNRAS} for {\it classical}\footnote{here we use the {\it classical scenario} to mean the merger of stellar mass black holes subsequent to the evolution of an isolated stellar field binary as described in \citep{Phinney1991, Tutukov1993,Belczynski2016}} stellar-mass BH merger to investigate the angular momentum evolution of the WR star. We refer the reader to that paper for background information. Essential elements of that work are reproduced here for ease of exposition, but detailed descriptions and derivations can be found there for the interested reader.

We will show that under very general assumptions in the classical scenario one expects a bimodal distribution for the effective spin parameter measured by LIGO. In section \ref{sec:a range} we review our basic set-up, in \ref{sec:expectations}
we describe why these assumptions lead to a few simple expectations regarding the effective spin parameter distribution. In  \ref{sec:discussion} we comment on the published measurements and conclude. 

\section{Angular Momentum Evolution of the WR star in a `Classic' Scenario}
\label{sec:a range}

The {\it classical scenario} remains one of the likely channels for forming the binary systems observed by LIGO \citep{LIGO2016}. In this picture, prior to the collapse of the second BH, the progenitor system is an isolated stellar field binary that is composed of a Wolf--Rayet (WR) star with a mass $M$ and the primary BH with a mass $M/q$, $q$ being defined as the mass ratio. In this brief note we follow the set up described in \citet{Kushnir2016MNRAS} to investigate the angular momentum evolution of the WR star.  LIGO observations provide a constraint on the component of the angular momentum in the direction of the orbital angular momentum through the measured parameter $\chieff$: 
\begin{equation}\label{eq:chieff}
\chieff=\frac{M_{1}\vec{a}_{1}+M_{2}\vec{a}_{2}}{M_{1}+M_{2}}\cdot\hat{L}
\end{equation}
(where $\vec{a}_{1}$ and $\vec{a}_{2}$ are the dimensionless BH spins, $\vec{a}= c\vec{S}/GM^{2}$, and $\hat{L}$ is the direction of orbital angular momentum).   $\chieff$ is constrained to the range $-1\le \chieff \le 1$. This component of the angular momentum is the most relevant for scenarios in which the spin is generated by tides or mass transfer in a binary system and thus will provide an important probe of these formation mechanisms. We assume that the final angular momentum of the star provides a good estimate for the angular momentum of the secondary BH. In what follows we will ignore corrections due to mass and angular momentum loss during the explosion that result in the formation of the BH. These are expected to be small. 

The angular momentum evolution of the WR star is determined by two competing process. Stellar winds decrease the angular momentum of the star while torque applied to the star by the primary BH increases the angular momentum, driving it towards synchronization. We now investigate the impact of each of these processes on the total angular momentum evolution.

For an initial orbital semi major axis, $d$, we can normalize the dimensionless spin of the star, $a$, to the orbital angular velocity, $\omega=\sqrt{G(M+M/q)/d^{3}}$ :
\begin{eqnarray}\label{eq:a}
a&=&\frac{cJ}{GM^{2}}=\frac{cr_{g}^{2} R^{2}}{GM}\left(\frac{1+q}{2q}\right)^{1/2}\left(\frac{2GM}{d^{3}}\right)^{1/2}\frac{\Omega}{\omega} \nonumber\\
&\equiv&\async\frac{\Omega}{\omega},
\end{eqnarray}
where $r_{g}^{2}$ is the (dimensionless) radius of gyration of the star related to the moment of inertia by $I=r_{g}^{2} R^{2}M$ and $\Omega$ is the angular spin velocity of the star. 
For synchronization between the stellar spin and the orbit ($\Omega=\omega$), we define $a=\async$. 

The torque $(\tau)$ applied to a star in a binary system was first calculated by \citet{Zahn1975}. We use the expression for the torque described in \citet{Kushnir:2016b} which 
leads to a simple dynamical equation for the spin: 
\begin{equation}\label{eq:a due to torque}
\dot{a}=\frac{c}{GM^{2}}\tau\equiv\frac{\async}{\ttau}\left|1-\frac{a}{\async}\right|^{8/3}. 
\end{equation}
The relevant time-scale depends on the profile of the star, but can be approximated by:
\begin{eqnarray}\label{eq:torque time2}
\ttau\approx10^{7}q^{-1/8}\left(\frac{1+q}{2q}\right)^{31/24}\left(\frac{\tmerge}{1\,\textrm{Gyr}}\right)^{17/8}\,\textrm{yr}.
\end{eqnarray}
We have introduced the merger time due to gravitational wave emission, $t_{\rm merge}$ which depends in a simple way on the initial orbital properties and component masses \citep{Peters1964}:
\begin{equation}\label{eq:tmerger}
\tmerge=\frac{5}{512}\frac{c^{5}}{G^{3}M^{3}}\frac{2q^{2}}{1+q}d^{4}. 
\end{equation}
It is important to note that $\ttau$ is a very strong function of distance, $\ttau \propto d^{17/2}$. 

Competing against the tidal torque is the loss of angular momentum resulting from mass-loss. 
Mass-loss from WR stars is complicated theoretically and observationally, thus the estimated rates are highly uncertain.  As in \citet{Kushnir:2016b} we incorporate this effect by modifying the equation of $a$ as:
\begin{equation}\label{eq:a evolution}
\dot{a}=\frac{\async}{\ttau}\left(1-\frac{a}{\async}\right)^{8/3}-\frac{a}{\twind},
\end{equation}
only valid for $a < \async$. We will take $\twind$ in the range $10^5$ to $10^6$ yrs to be a reasonable estimate but we note that our conclusions are robust to this choice. We will integrate this equation until the end of the life of the WR star, which we take to be \ $t_{WR}\approx 3\times 10^5$ yrs.  

\section{Simple Expectations}
\label{sec:expectations}

In this section we will describe some of the consequences of the simple formulae described above. 

\paragraph*{Timescales:}  There are two important distance scales in the problem. The first one, $\dmerge$, will be defined as the distance that corresponds to a merger time equal to a Hubble time $t_{H}$, {\it ie.} $\tmerge \sim 10^{10}$ yrs. For separations larger than this, the binary does not have enough time to merge in the lifetime of the universe. The other important distance, $\dtau$, will be defined as the distance that corresponds to $\ttau \sim \twr$ (the Wolf-Rayet lifetime),  inside of which torques are important and synchronize the binary and outside of which torques are irrelevant.   

\begin{figure}
\includegraphics[width=0.5\textwidth]{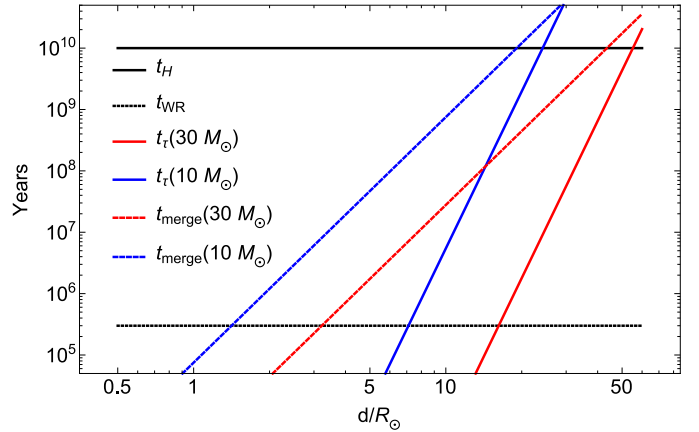}
\caption{Timescales as a function of separation for binaries composed of a WR and a BH of equal mass. Two examples are shown $M=10\ M_\odot$ and  $M=30\ M_\odot$.
\label{fig:timescales}}
\end{figure}

There are two points to note: first although $\dtau < \dmerge$ they are of comparable magnitude. We give some representative numbers in Table \ref{table1} for systems of different masses. The second point, evident in Figure \ref{fig:timescales} is that
both $\tmerge$ and $\ttau$ are very strong functions of separation ($\tmerge \propto d^{4}$ and  $\ttau\propto d^{17/2}$). In particular this implies that the transition between $t_\tau << t_{WR}$ and $t_\tau >> t_{WR}$ occurs over a very small range of separations. Furthermore, because of the steep dependence, the fact that the locations where tidal torques are important is comparable to the distance required for the binary to merge in a Hubble time is relatively insensitive to the details.   

\begin{table}
\begin{tabular}{|l|l|l|ll}
\hline
M  & $\dtau$ & $\dmerge$     \\ \hline
10 $M_\odot$ & 7  $R_\odot$ & 19   $R_\odot$  \\ \hline
20 $M_\odot$ & 12  $R_\odot$ & 32  $R_\odot$   \\ \hline
30 $M_\odot$ & 16 $R_\odot$  & 44  $R_\odot$  \\ \hline
\end{tabular}
\centering
\caption{Separations corresponding to $\ttau = 3 \times 10^{5}$ yrs ($\dtau$) and $\tmerge = 10^{10} $ yrs ($\dmerge$) for binaries composed of a WR and a BH of equal mass ($M$).  }
\label{table1}
\end{table}

\paragraph*{Possible separations} 

Let us assume that the distance between the WR star and the BH is set by the details of late-stage massive star evolution.   In that case neither $\dmerge$ nor $\dtau$ are distances relevant for the dynamics as the stellar evolution happens on very short timescales compared to $\ttau$ and $\tmerge$.  In all examples in Table \ref{table1} $\dmerge$ and $\dtau$ are within a factor of three of each other.  Given how close those distances are, one might expect a similar rate of occurrence of binaries on both sides of the line given by  $d=\dtau$.

\begin{figure}
\includegraphics[width=0.5\textwidth]{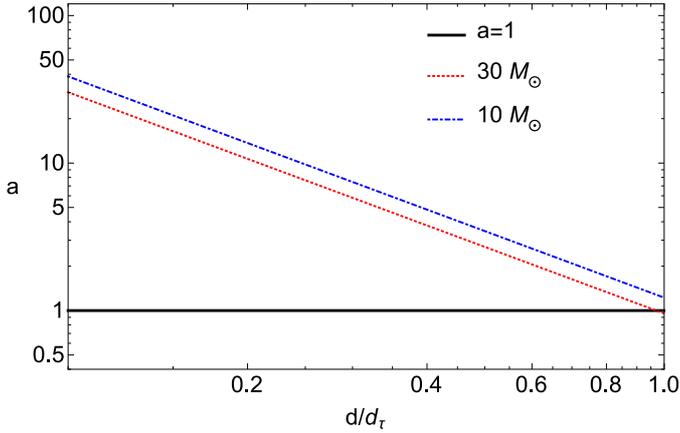}
\caption{Spin parameter of the WR star at synchronized rotation  ($\async$)   as a function of separation for binaries  of equal masses $M=10\ M_\odot$ or $M=30\ M_\odot$. It follows a simple scaling, $\async \propto d^{-3/2}$ (equation \ref{eq:a}). In making this figure we used $r_g=0.075$ and  $\log_{10}(R/R_{\odot})=-0.70+0.71\log_{10}(M/M_{\odot})$ \citep{Kushnir2016MNRAS}.
\label{fig:syncspin}}
\end{figure}

\paragraph*{Spin when tidally locked:} There is another interesting point which is illustrated by Figure \ref{fig:syncspin}. Over most of the range where tides are important, a tidally locked WR star would have angular momentum in excess of that of a maximally rotating BH of that mass. Thus one would expect that any WR-BH binary with separations $d<\dtau$ would lead to a maximally spinning secondary black hole. 

\paragraph*{Spin at large separations:} When the separation of the binary is $d > \dtau$ the final spin of the WR before collapse $a(\twr)$ just reflects the initial spin suppressed by the losses induced due to the wind or any other angular momentum loss-mechanism. In our simple model $a(\twr, d >> \dtau) = e^{-\twr/\twind} a_{initial}$. If this already corresponds to $a>>1$ then both at large and small binary separations the secondary BH will have nearly maximal spin. At large separations the distribution of spins will be a direct result of the initial conditions while at small separations the spin of the secondary BH will be aligned with the orbital spin and be close to maximal. 
Observations by LIGO already imply that not all BH are rapidly spinning at birth (at least in the direction of the angular momentum of the orbit) so in what follows we will assume that the spin at birth is small compared to one. In that limit our conclusions are not sensitive to the distribution of initial spins. 

\begin{figure}
\includegraphics[width=0.5\textwidth]{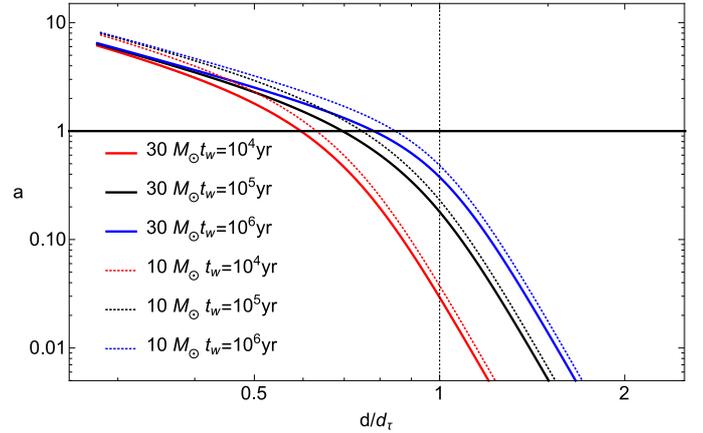}
\caption{Spin at the end of the lifetime of the WR star as a function of separation $a(\twr, d)$ for binaries with equal mass WR and BH of either $M=10\ M_\odot$ or $M=30\ M_\odot$. The initial condition of the spin was set to $a_{initial} =0$. We show two representative examples for $\twind = 10^5, \ 10^6$ yrs and perhaps a more extreme choice $\twind = 10^4$ yrs.  
\label{fig:finalspin}}
\end{figure}

To illustrate the key effects we solve equation (\ref{eq:a evolution}) for a few representative examples.  { We assume that in the absence of tidal torquing, the spin of the WR is sufficiently slow that only dynamical (rather than natal) effects are ``in play"}. To illustrate the dependence on other parameters we will fix $a_{initial} =0$. This choice will make the spin go to zero at large separations, if the initial birth spin is non-zero then the result would asymptote to  $e^{-\twr/\twind} a_{initial}$. We show $a(\twr, d)$ in Figure  \ref{fig:finalspin} where we show the separation in units of $\dtau$. When the final spin exceeds $a=1$ we will assume that the BH that forms will have maximal spin.  The two cases shown, $M=10\ M_\odot$ and $M=30\ M_\odot$ are very similar, with relatively fast transition between a non-spinning secondary BH and a maximally rotating secondary. Although the exact location of the transition depends on $\twind$ the dependence is rather minor, reflecting the steep dependence of $\ttau$ with separation. 
In the region of parameters where $\twind << \ttau$ at late times the system will equilibrate to $\async/(1+\ttau/\twind)$. Because we are taking $\twind$ and $\twr$ to be similar and we define $\dtau$ as the place where $\ttau=\twr$ this is close to $\async$. Quickly, as one moves to smaller $d$, $\ttau << \twind$ so  $\async/(1+\ttau/\twind) \rightarrow \async$.  For the cases considered here, this effect only changes the details of the transition region but if we chose a significantly smaller $\twind$ it will make some difference over a moderate range of $d$ { as the longer the wind operates over the lifetime of the WR, the larger the probability of finding a slowly spinning secondary BH. }

\section{Discussion}
\label{sec:discussion}

The main conclusion of this note is that in the classical isolated field-binary scenario the spin distribution of the secondary BH should be bimodal. At small separations tidal torques spin up the secondary BH to the maximal value. Due to the steep dependence with separation, the transition is rather abrupt as a function of separation. If the initial spins are small then one expects the gravitational wave sources to have a bimodal distribution of effective spins.  { In this case, the fraction of secondary BH with maximal spin should be comparable to those spinning slowly. } 

\begin{figure}
\includegraphics[width=0.5\textwidth]{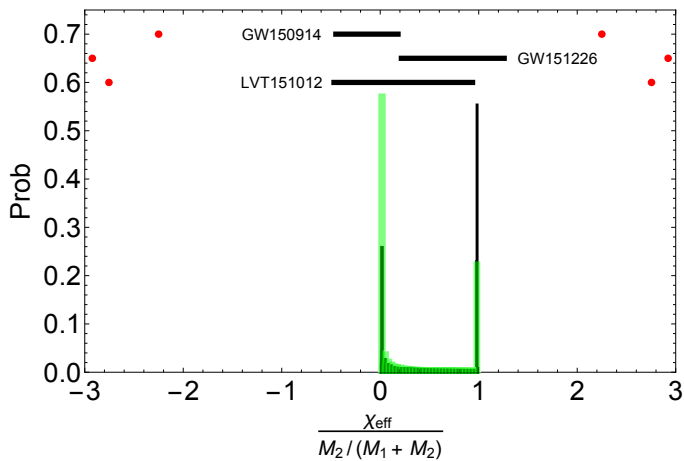}
\caption{Distributions of $\chieff$ for a pair of $M=30\ M_\odot$ BHs under the assumption that the birth spins are negligible and that the initial separation is uniform (green) or uniform in log-separation (black) with a minimum separation given by the radius of the star and and a maximum given by $\dmerge$. 
$\chi_{eff}(M_1+M_2)/M_2$ is constrained to be between zero and one in the scenario where the primary BH has negligible spin. This range was divided into 30 equal size bins and the expected fraction of systems in each bin is shown. The results for a pair of $M=10\ M_\odot$ BHs are very similar and not shown here. We also show the measurements from the reported LIGO events. We take the measured $\chieff$ and mass ratio $q$ with their reported 90\% credible intervals  to compute the allowed range for $\chieff (M_1+M_2)/M_2$. For simplicity we added both errors in quadrature. The possible range of observed values for $\chieff (M_1+M_2)/M_2$ is indicated by red dots for each event.  This range is bounded by the case both BHs are maximally spinning in the same direction.
\label{fig:distribution}}
\end{figure}

As an illustration we consider the case in which the birth spins are small and the initial separations of the binary are distributed either uniformly in separation or uniformly in log-separation. The resulting distributions are shown in Figure \ref{fig:distribution}.  We have normalized $\chieff$ by the predicted maximal value in this scenario -- $M_{2}/(M_{1}+M_{2})$, achieved when the secondary BH is maximally spinning but the primary has negligible spin. The bimodal nature of the distribution is readily apparent. { Of course, the details of the distribution depend on the distribution of the initial separations, the initial distribution of natal spins (assumed as negligible in Figure~\ref{fig:distribution}), the mass ratio, spin-orbit alignment, wind efficiency, etc. But the robust conclusion is that the fraction of systems with maximally spinning BH secondaries will be similar to that with slowly spinning secondaries, leading to two clear peaks in the $\chieff$ distribution.}

Note that if the primary black hole is maximally spinning while the secondary has negligible spin the value of  $\chieff/(M_{2}/(M_{1}+M_{2}))$  would be $1/q$. If both BH have maximal spin we would have $\chieff/(M_{2}/(M_{1}+M_{2}))= (1+q)/q$, a value we illustrate with red dots in figure \ref{fig:distribution}.

Let us now discuss the published LIGO events \citep{LIGO2016} in the context of what we have derived in the previous section. 

\paragraph*{GW150914} The first LIGO event was characterized by $M_1/M_\odot = 36.2^{+ 5.2}_{-3.8}$ and  $M_2/M_\odot = 29.1^{+ 3.7}_{-4.4}$. No spin in the orbital-spin direction was detected,  $\chieff =0.06^{+0.14}_{-0.14}$. For this system, $\dmerge \sim 40\ R_\odot$ and $\dtau \sim 15\ R_\odot$. If we take the lack of spin to mean that the separation is larger than $\dtau$ we conclude that $ 15\ R_\odot < d < 40\ R_\odot$. The minimum separation also implies a minimum $\tmerge > 1 \times 10^8$yrs,  already reported in \citep{Kushnir2016MNRAS}.  The lack of spin also implies that the WR descendent BH was not spinning fast at birth and that the primary BH was not maximally spinning.

\paragraph*{GW151226} The second LIGO event was characterized by $M_1/M_\odot = 14.2^{+ 8.3}_{-3.7}$ and $M_2/M_\odot = 7.5^{+ 2.3}_{-2.3}$. In this case a spin in the direction of the orbit was clearly detected $\chieff =0.21^{+0.2}_{-0.1}$.  For this system $\dmerge \sim 17\ R_\odot$ and $\dtau \sim 7\ R_\odot$. If we assume that the measured spin reflects the spin of the secondary BH acquired as we have discussed, we can conclude  that in this case $ d < 7\ R_\odot$ which would imply that $\tmerge < 2 \times 10^8$yrs. In this case the secondary black hole would be maximally spinning which would lead to a predicted $\chieff \sim 7 / (14+7) \sim 0.33$, which is consistent with the observed value. { We have assumed that the lowest mass BH is the secondary BH, otherwise $\chieff \sim 14 / (14+7) \sim 0.66$ which is in tension with the data.  Indeed, as in GW150914, the primary BH cannot be close to maximally spinning in this picture (modulo severe spin-orbit mis-alignment).}

\paragraph*{LVT151012} This event is considered less significant so less weight should be given to it, however we compute the relevant scales for completeness.  This system is characterized by $M_1/M_\odot = 23.2^{+ 18}_{-6}$ and  $M_2/M_\odot = 13^{+ 4}_{-5}$ and, as in GW150914, no spin was detected,  $\chieff =0.0^{+0.3}_{-0.2}$. For this system $\dmerge \sim 19\ R_\odot$ and $\dtau \sim 7\ R_\odot$. {These results are consistent with having both BHs with negligible spin but inconsistent with both BHs maximally spinning.  They are also inconsistent with a case where the heaviest black hole is maximally spinning, although the data is marginally consistent with the smallest mass black hole being maximally spinning.}  At face value, if none of the BHs are maximally spinning we infer for this system $\tmerge > 2 \times 10^8$yrs.

\paragraph*{Other BH Spins} {In the isolated field binary scenario, we have established that there is a sharp transition in orbital separation such that for small separations the secondary black hole is maximally spinning while otherwise it just retains the memory of its initial spin (modified by the loss of angular momentum due to the wind). }This appears consistent with the LIGO data so far assuming that the initial spin of the WR star is small and if the primary black hole is also not spinning rapidly.  {How do we reconcile these conclusions with the reported (high) spins of High-mass X-ray Binaries (HMXBs), thought to be a snapshot of the pre-merger phase of this scenario?}  Indeed, the HMXBs such as Cyg X-1, LMC X-1 and M33 X-7 all have high and precisely measured spins, however, there remains some uncertainty in these measurements regarding the spin-orbit alignment \citep{McClintock2014}. As we have shown, a primary BH that is maximally spinning in the direction of the orbit is clearly in contradiction with the first LIGO event where no spin was detected. 

The expectations of our simple model as well as the existing measurements of spins in HMXBs make it very interesting to use LIGO data to determine if the BH have maximal spin.  Existing data is already able to shed some light on this question.  As more data from LIGO becomes available, our simple expectation about distribution of $\chieff$ will be easily tested and thereby shed light on the formation the channel for the sources of gravitational waves.

\section*{Acknowledgements}
MZ is supported by the NSF grants PHY-1213563, AST-1409709, and PHY-1521097. DK is supported in part by the Benoziyo fund for the advancement of Science.  JAK thanks the organizers and participants of the KITP program ``Black Holes from LIGO", specifically Sterl Phinney, for interesting discussions.  MZ thanks Morgan MacLeod and Dong Lai. 









\end{document}